\documentclass[acus]{JAC2003}

\usepackage{graphicx}
\usepackage{booktabs}
\usepackage{subfig}
\usepackage{array}
\usepackage{upgreek}
\usepackage{amsmath}

\newcommand{\bA}{\mathbf{A}}
\newcommand{\bv}{\mathbf{v}}

\begin{document}
\title{Theoretical Field Limits for Multi-Layer Superconductors\thanks{Work supported by DOE award number DE-SC0002329 and in part by the EU under REA grant agreement CIG-618258.}}

\author{Sam Posen$^1$\thanks{sep93@cornell.edu}, Gianluigi Catelani$^2$, Matthias U. Liepe$^1$, James P.~Sethna$^3$, and Mark K. Transtrum$^4$\\
$^1$Cornell Laboratory for Accelerator-Based Sciences and Education, Ithaca, NY\\
$^2$Forschungszentrum J{\"u}lich Peter Gr{\"u}nberg Institut (PGI-2), J{\"u}lich, Germany\\
$^3$LASSP, Physics Department, Clark Hall, Cornell University, Ithaca, NY\\
$^4$ Department of Bioinformatics and Computational Biology,\\
University of Texas M. D. Anderson Cancer Center, Houston, Texas}

\maketitle

\begin{abstract}
The SIS structure---a thin superconducting film on a bulk superconductor separated by a thin insulating film---was propsed as a method to protect alternative SRF materials from flux penetration by enhancing the first critical field $B_{c1}$. In this work, we show that in fact $B_{c1}$ = 0 for a SIS structure. We calculate the superheating field $B_{sh}$, and we show that it can be enhanced slightly using the SIS structure, but only for a small range of film thicknesses and only if the film and the bulk are different materials. We also show that using a multilayer instead of a single thick layer is detrimental, as this decreases $B_{sh}$ of the film. We calculate the dissipation due to vortex penetration above the $B_{sh}$ of the film, and find that it is unmanageable for SRF applications. However, we find that if a gradient in the phase of the order parameter is introduced, SIS structures may be able to shield large DC and low frequency fields. We argue that the SIS structure is not beneficial for SRF cavities, but due to recent experiments showing low-surface-resistance performance above $B_{c1}$ in cavities made of superconductors with small coherence lengths, we argue that enhancement of $B_{c1}$ is not necessary, and that bulk films of alternative materials show great promise.
\end{abstract}

\section{Introduction}
\label{sec:Intro}
SRF researchers have begun a significant effort to develop alternative materials to niobium, superconductors that could offer higher accelerating gradients $E_{acc}$ and/or lower surface resistances $R_s$ at a given temperature. There are several promising candidates, but most of them suffer from two potential liabilities. First, as shown in Table \ref{tab:mats}, they have relatively small first critical fields $B_{c1}$, the magnetic field at which it becomes energetically favorable for a vortex to be inside the superconductor. Second, they have relatively small coherence lengths $\xi$. Vortex penetration is prevented at fields significantly above $B_{c1}$ by an energy barrier, but surface defects on the order of $\xi$ can reduce this barrier. These materials have $\xi$ on the order of a few nm, compared to tens of nm for niobium, making even very small defects a potential vulnerability. As a result, there has been significant concern in the SRF community over whether vortex dissipation will occur if these materials are exposed to fields that bring them into the metastable state between $B_{c1}$ and $B_{sh}$, the superheating field at which the energy barrier is reduced to zero for an ideal surface.

\begin{table}[htbp]
\begin{center}
\begin{tabular}{| c | r | r | r | r |}
\hline Material & $\lambda$ [nm] & $\xi$ [nm] & $B_{c1}$ [T] & $B_{sh}$ [T] \\ \hline\hline
    Nb & 40 & 27 & 0.13 & 0.24 \\ \hline
    Nb$_3$Sn & 111 & 4.2 & 0.042 & 0.36 \\ \hline
    NbN & 375 & 2.9 & 0.006 & 0.15\\ \hline
    MgB$_2$ & 185 & 4.9 & 0.017 & 0.19 \\ \hline
\end{tabular}
\caption{Material properties of niobium and three promising alternative SRF materials. The penetration depth $\lambda$ is calculated using Eqn 3.131 in \cite{Tinkham}. The correlation length $\xi$ is calculated using the equations in \cite{Orlando}. For Nb a RRR of 100 was assumed. For MgB$_2$, $\lambda$ and $\xi$ are not calculated, as the experimental values are given in the reference. For calculations, $B_c=\phi_0/(2\sqrt{2}\pi\xi\lambda)$ is used, where $\phi_0$ is the flux quantum \cite{Tinkham}. $B_{c1}$ for Nb found from power law fit to numerically computed data from \cite{Hein} and for strongly type II materials is found from Eqn 5.18 in \cite{Tinkham}. $B_{sh}$ for Nb is found from \cite{Dolgert} and for others calculated from $B_c\sqrt{20}/6$ (valid only for strongly type II materials near $T_c$) \cite{Transtrum}. Nb data from \cite{Maxfield}, Nb$_3$Sn data from \cite{Hein}, NbN data from \cite{Oates}, and MgB$_2$ data from~\cite{Wang}. Note that the two gap nature of MgB$_2$ may require more careful analysis than is performed here.}
\label{tab:mats}
\end{center}
\end{table}

A. Gurevich proposed \cite{Gurevich} a method to avoid the potentially vulnerable metastable state altogether. Pointing to the enhancement of parallel $B_{c1}$ in films with thickness $d$ smaller than the penetration depth $\lambda$, he suggested coating a niobium cavity with alternating layers of insulator (I) and thin film superconductor (S). With such a SIS structure, he proposed it might be possible to take advantage of the high $B_{sh}$ and low $R_s$ of the alternative superconductors used in the thin films without the disadvantage of their small $B_{c1}$. SRF researchers have been putting significant effort into developing SIS multilayers, and they are producing excellent work \cite{Antoine}\cite{Tajima}\cite{Beringer}\cite{VF}\cite{xi}\cite{Proslier}\cite{MITOates}.

In this work, we will start by showing that in practice there is no enhancement of $B_{c1}$ for SIS films---we will show that in fact $B_{c1}$ is zero for such a structure. Next we will study the superheating field of SIS structures and show that for a homolaminate, $B_{sh}$ is always lower than the bulk value, and for a heterolaminate, only a small increase in $B_{sh}$ is possible. Following this, we show that SIS structures cannot be used above $B_{sh}$ for SRF applications, as the heating would be unmanageable. We then consider SIS multilayer films for DC and at low frequencies, and show that they may be effective in screening large fields by setting up a gradient in the phase of the order parameter. Finally, we consider the outlook for alternative materials for SRF cavities and show that recent developments give reason for strong optimism for bulk films.

\section{No Bc1 Enhancment}
\label{sec:Bc1}
Tinkham \cite{Tinkham} defines $B_{c1}$ as the field at which ``the Gibbs free energy [has] the same value whether the first vortex is in or out of the sample.'' For a SIS, the sample under consideration should be the full structure \cite{footgibbs}. Stejic et al. \cite{Stejic} calculate the Gibbs free energy of a vortex in a thin film superconductor immersed in a parallel external field. They show that $B_{c1}$ of the film is enhanced relative to the bulk value, according to 
\begin{equation}
B_{c1}=\frac{2\phi_0}{\pi d^2}\left(\ln{\frac{d}{\xi}+\gamma}\right)
\label{eq:bc1thin}
\end{equation}
where $\phi_0$ is the flux quantum, $\gamma=-0.07$ and $d<<\lambda$. However, if a SIS structure is used to screen Nb SRF cavities, the geometry is quite different than that of an isolated film. How does Stejic's expression for $B_{c1}$ change when the film is screening a bulk superconductor? In this case, it will have a B-field gradient across it, which will affect the free energy. We can use the same formalism as Stejic to calculate the Gibbs free energy for this case, and use it to find $B_{c1}$ and $B_{sh}$\cite{foot2}.

\begin{figure}[htbp]
\begin{center}
\includegraphics[width=0.48\textwidth,angle=0]{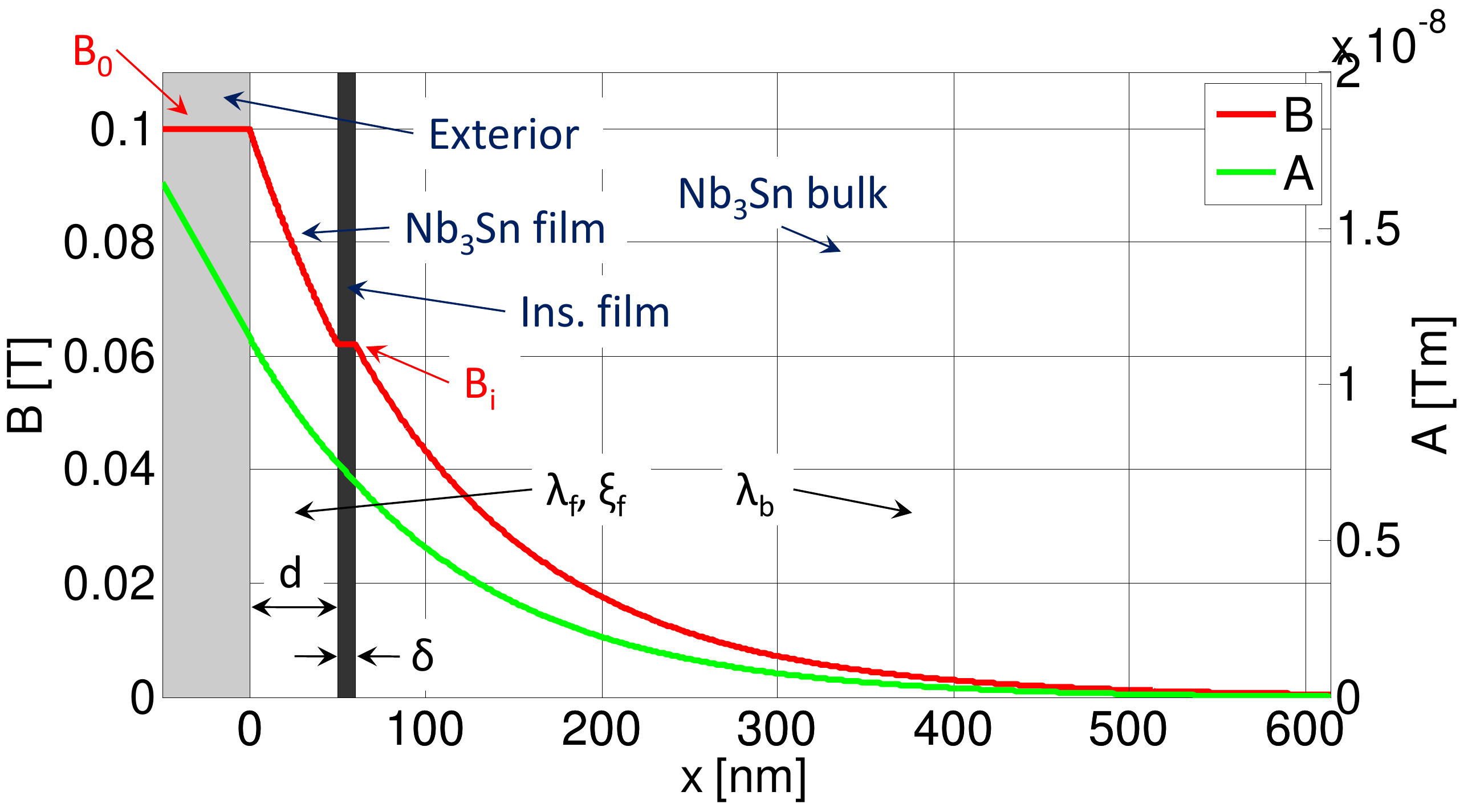}
\end{center}
\caption{Geometry of the structure under consideration. The amplitudes of the magnetic field and the vector potential are plotted as a function of distance into the structure.}
\label{fig:geometry}
\end{figure}

Consider a single layer SIS structure, as shown in Fig.~\ref{fig:geometry}. A strongly type II superconducting film of thickness $d$, penetration depth $\lambda_f$, and coherence length $\xi_f$ is separated from a bulk superconductor with with penetration depth $\lambda_b$ by an insulating film of thickness $\delta$. The superconducting film is screening the bulk from a parallel magnetic field with amplitude $B_0$. The screened field inside the bulk region has amplitude $B_i$. In our geometry, the x-axis is perpendicular to the film, pointing into it, with origin at the interface with the exterior. The z-axis is aligned with the magnetic field.

Stejic shows that the Gibbs free energy of a vortex in a superconductor can be determined from the value of two magnetic fields evaluated at the vortex location $r_0$: the Meissner-screened external field $B_M$ and the field generated by the vortex in the film $B_V$.

\begin{equation}
\mathcal{G}=\frac{\phi_0}{\mu_0}\left(B_V(r_0)/2+B_M(r_0)\right)
\label{eq:gibbs}
\end{equation}

$B_M$ can be found by minimizing the free energy in the structure when no vortex is present. This procedure gives:

\begin{equation}
B_M=\frac{B_0+B_i}{2}\frac{\cosh{\frac{x}{\lambda_f}}}{\cosh{\frac{d}{2\lambda_f}}}-\frac{B_0-B_i}{2}\frac{\sinh{\frac{x}{\lambda_f}}}{\sinh{\frac{d}{2\lambda_f}}}
\label{eq:Bm}
\end{equation}

where $B_i$ is given by

\begin{equation}
B_i=B_0\left[\frac{\delta+\lambda_b}{\lambda_f}\sinh{\frac{d}{\lambda_f}}+\cosh{\frac{d}{\lambda_f}}\right]^{ -1}
\label{eq:Bi}
\end{equation}

Stejic gives a relatively simple expression for $B_V$ for the case when $d<<\lambda$, but this would restrict us to very thin films. To study the full range of thicknesses, we turn to the more general expression from Shmidt \cite{Shmidt} (this expression assumes $r_0=(x_0,0)$), which agrees with Stejic's expression for very small films:

\begin{equation}
B_V=\frac{2\phi_0}{\lambda^2d}\sum\limits_{n=1}^\infty\int\limits_{-\infty}^\infty \frac{dk}{2\pi}e^{iky}\frac{\sin(\pi n x/d)\sin(\pi n x_0/d)}{k^2+(\pi n x_0/d)^2+1/\lambda^2}
\label{eq:BV}
\end{equation}

\begin{figure}[tbp]
\begin{center}
\includegraphics[width=0.48\textwidth,angle=0]{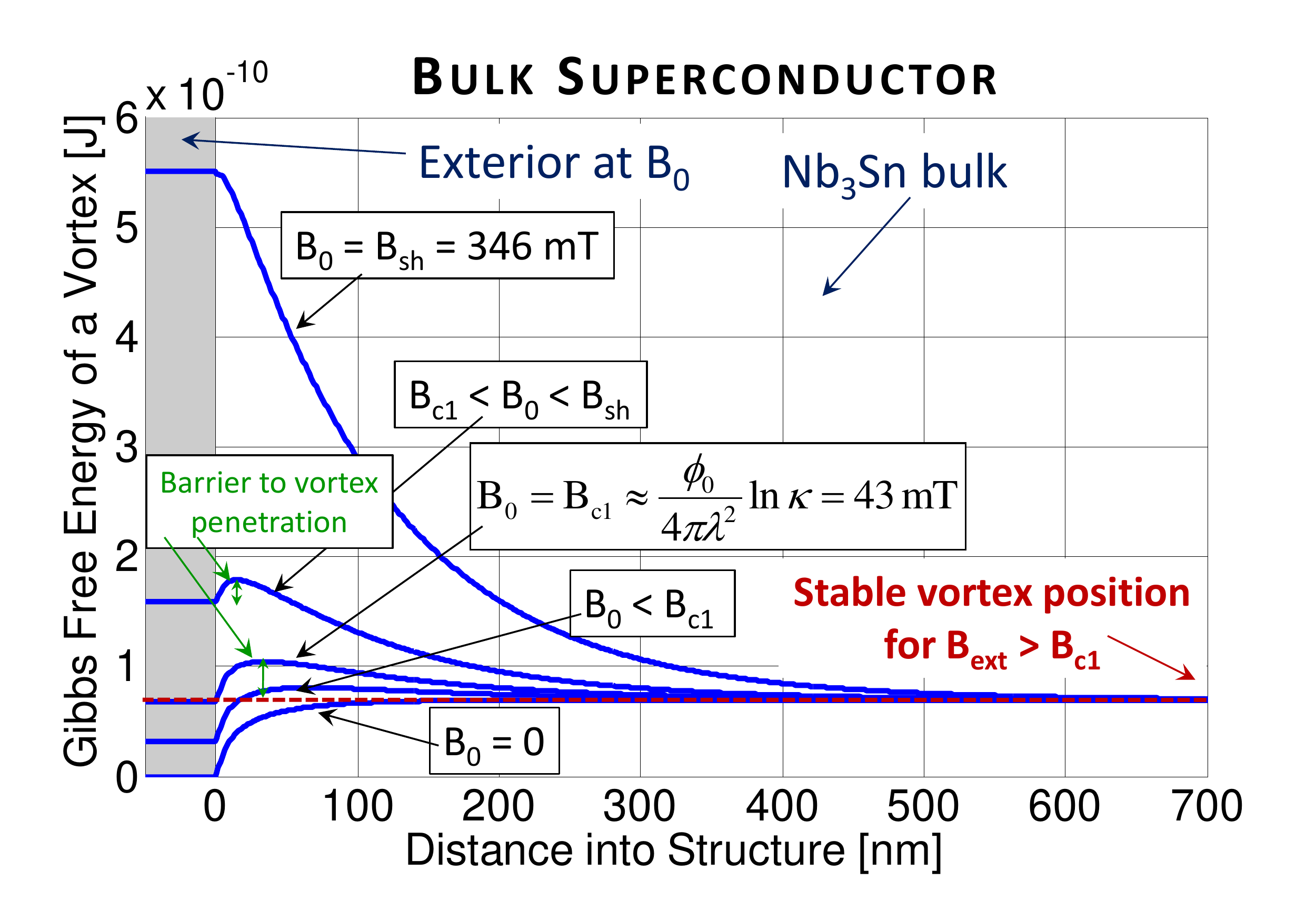}
\includegraphics[width=0.48\textwidth,angle=0]{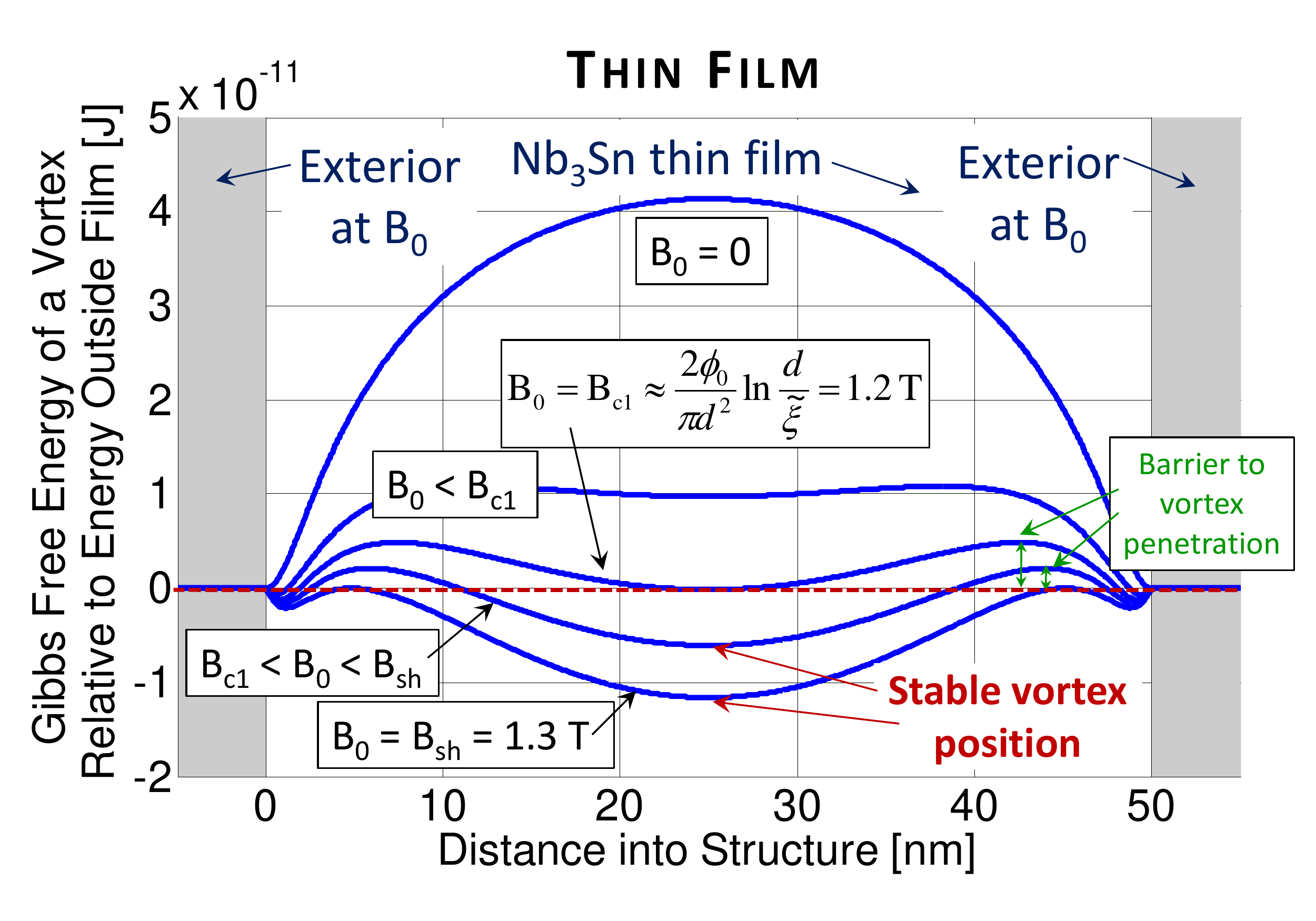}
\includegraphics[width=0.48\textwidth,angle=0]{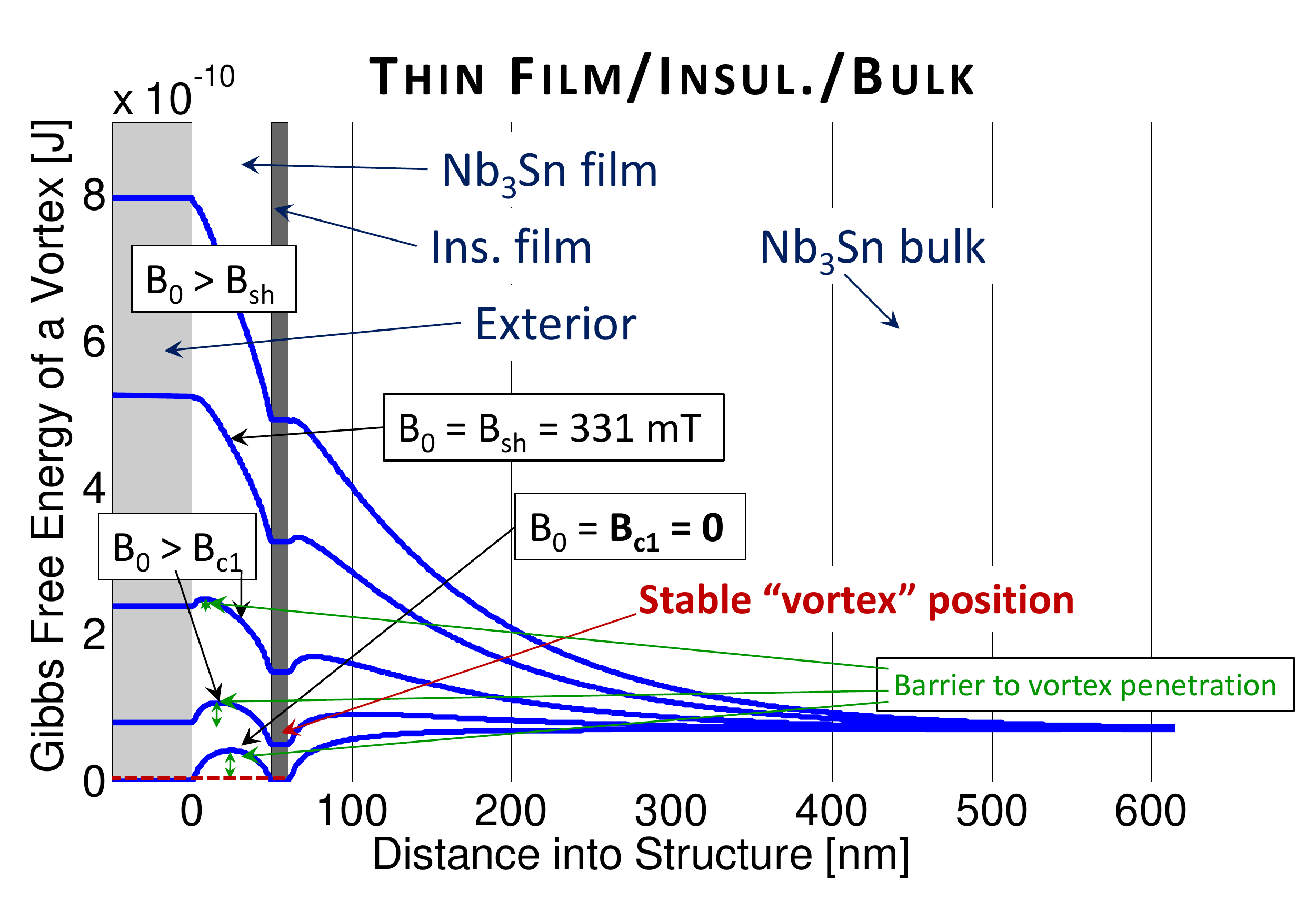}
\end{center}
\caption{Gibbs free energy at various fields for a single vortex in (top) bulk Nb$_3$Sn, (center) a 50 nm Nb$_3$Sn thin film, and (bottom) a SIS structure with a 50 nm Nb$_3$Sn film on a Nb$_3$Sn bulk. $B_{c1}$ is the smallest field at which there is a position inside the structure where the free energy for a vortex is smaller than the value outside. $B_{sh}$ is the field at which the energy barrier to vortex penetration disappears. The top and center plots show the $B_{c1}$ enhancement for a thin film compared to a bulk. The bottom plot shows that for a SIS structure $B_{c1}=0$. The expression for the thin film $B_{c1}$ is not valid for the SIS structure because it assumes that the first stable vortex position will be at the center of the film. However, for the SIS structure, the first stable vortex position occurs on the side of the film adjacent to the insulating layer.} 
\label{fig:Gplots}
\end{figure}

We can check our procedure by choosing $d>>\lambda$, such that the film behaves as a bulk supercondcutor. This calculation is shown in the top plot of Fig.~\ref{fig:Gplots}. $B=B_{c1}$ when the free energy outside the superconductor is equal to that when a vortex is deep in the bulk. $B=B_{sh}$ when the barrier to flux penetration is reduced to zero (this plot is very similar to the one from Bean and Livingston's 1963 paper \cite{BL}).

We can study a single thin film (not in a SIS structure) by setting $B_i=0$ in Eqn.~\ref{eq:Bm}. This calculation is shown in the center plot of Fig.~\ref{fig:Gplots} (the free energy outside the film is subracted from each of the plots for clarity). In this case, there is no bulk, so the first location at which the free energy drops below the external value at high fields is in the center of the film. This would be the stable position for a single vortex above $B_{c1}$. Both $B_{c1}$ and $B_{sh}$ are much higher for the film than the bulk.

Finally, we plot the free energy of vortex in a single SIS structure in the bottom plot of Fig.~\ref{fig:Gplots}. In contrast to the previous case, only one side of the thin film is exposed to the external magnetic field. The field at the other side is smaller due to screening by film. Since $B_V$ = 0 at the edges of the film, Eqn. \ref{eq:gibbs} shows that the free energy in the insulating layer is lower than the free energy outside. The film provides screening at any finite $B_0$ below the second critical field, so for $B_0>0$, the energetically favorable configuration is for flux to be trapped in the insulating layer. As we explain below, this implies that in practice for the SIS structure, $B_{c1}$ is zero.

Why is Eqn. \ref{eq:bc1thin} describing the enhancement of $B_{c1}$ in a lone thin film not applicable for the SIS structure? This expression assumes that the first stable vortex position will occur in the center of the film. It predicts when the free energy at the center of the film will dip below the value of the free energy in the exterior. However, for the SIS structure, the free energy at the insulator side of the film will dip below the exterior value at fields much smaller than this.

Fig. \ref{fig:Gplots} shows that at moderate fields, when $B_i$ is below $B_{c1}$ of the bulk superconductor, there is no stable position for a vortex in either the bulk or the film. In effect, both superconductors are below their individual $B_{c1}$, so it is not immediately obvious if $B_{c1}$ of the overall structure is important. Let us consider the implications of it being energetically favorable for a vortex to pass through the film, and have its flux trapped in the insulating layer. Once the flux is trapped in this way, it is non-dissipative under RF fields (unlike a vortex, which has a normal conducting core). However, as the vortex penetrates through the film to the insulator, dissipation occurs due to drag, and later we will show that this dissipation is too strong to be tolerable for SRF applications. Therefore it is the $B_{c1}$ of the SIS structure that is important, not that of the individual superconductors. Above $B_{c1}$, the structure is in a metastable state: only the energy barrier of the film prevents quench-inducing vortex penetration.

\section{The Superheating Field in Multilayers}
\label{sec:Bsh}

\begin{figure}[htbp]
\begin{center}
\includegraphics[width=0.48\textwidth,angle=0]{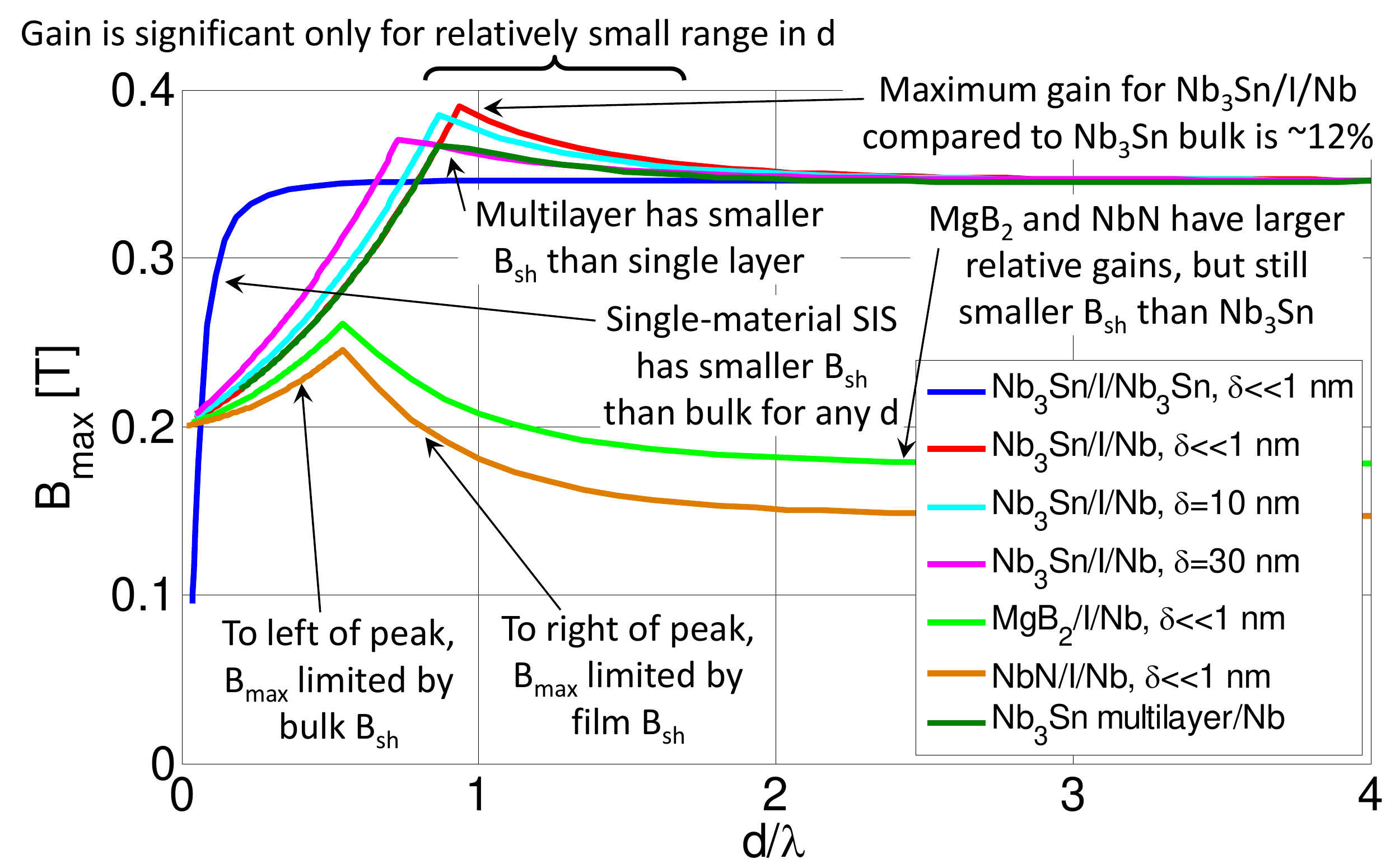}
\end{center}
\caption{Maximum field below $B_{sh}$ of both the film and the bulk as a function of film thickness for various film materials in a SIS structure with Nb. The effect of varying the insulator thickness $\delta$ is shown for the Nb$_3$Sn film, as is the effect of splitting the film thickness $d$ over 5 equally thick multilayers with thin separating insulators.}
\label{fig:BshvsdMulti}
\end{figure}

We have shown that the $B_{c1}$ of SIS structures is always zero. However, they may still offer a significant advantage over bulk films if they have a higher superheating field. In Fig.~\ref{fig:Gplots}, the barrier height is indicated for fields between $B_{c1}$ and $B_{sh}$. We can find the superheating field easily using the same free energy calculations to find the field at which the barrier is reduced to zero. In Fig.~\ref{fig:BshvsdMulti}, the maximum applied field at which both the film and the bulk are exposed to fields below their respective $B_{sh}$ is plotted as a function of film thickness. Various materials and insulator thicknesses are considered \cite{foot1}.

\begin{figure}[htbp]
\begin{center}
\includegraphics[width=0.48\textwidth,angle=0]{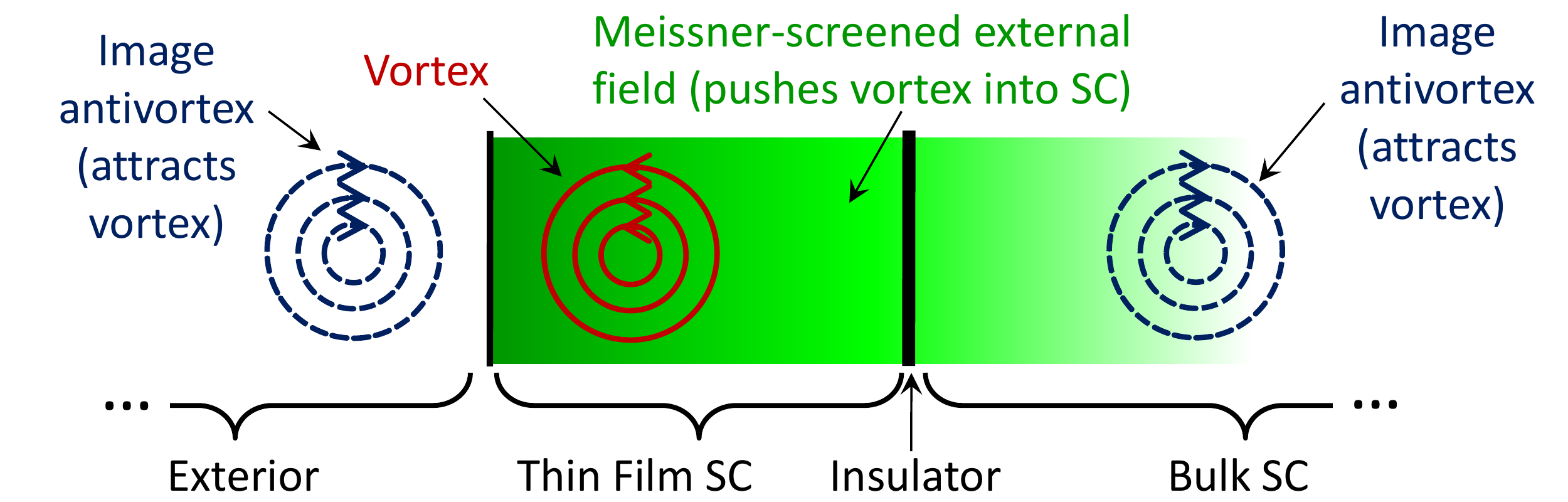}
\end{center}
\caption{Forces on a vortex in a homolaminate. As the film is made thinner, the second image antivortex has a stronger pull on the vortex, lowering the barrier to flux penetration.} 
\label{fig:vortex}
\end{figure}

Consider the curve for the Nb$_3$Sn/insulator/Nb$_3$Sn structure. Calculations show that for a homolaminate like this one---a homolaminate being a SIS structure in which the film is the same material as the substrate---the film always reaches its $B_{sh}$ before the bulk, and the thinner the film, the lower its $B_{sh}$. The highest $B_{sh}$ occurs when the film is so thick that it approximates a bulk superconductor. This can be understood in terms of the forces on a vortex (which can be derived from Eqn.~\ref{eq:gibbs}), as shown in Fig.~\ref{fig:vortex}. $B_M$, the Meissner-screened external field, pushes the vortex into the film due to the gradient in the field. The boundary condition imposed by $B_V$, the magnetic field of the vortex, can be satisfied by an image antivortex outside of the boundary, which creates a force that pulls the vortex out of the film \cite{BL}. Assuming that the insulating layers are very thin (this is optimal, as will be shown later), then as Eqn \ref{eq:Bi} shows, if $\lambda_b=\lambda_f$, then the force due to $B_M$ is the same no matter how thick the film is. However, there are two image antivortices contributing to $B_V$: one for the external boundary, and one for the boundary at the insulating layer. For a thin film, the image antivortex at the insulating layer will have a significant attractive force, pulling the vortex into the film. Since this is the only change in the forces on the vortex as the film thickness changes, a homolaminate will only have its $B_{sh}$ decrease as the film becomes significantly smaller than a bulk. Note that this same logic applies no matter how many times the superconductor is divided---a homolaminate SIS multilayer with many thin film layers will still experience a reduced $B_{sh}$.

Only one curve in Figure \ref{fig:BshvsdMulti} is a homolaminate; the rest are heterolaminates, in which the film material is different than the bulk. For a very thin heterolaminate, the film does not provide much screening for the bulk, and $B_i$ reaches the bulk's $B_{sh}$ before the thin film barrier disappears. A very thick film behaves as a bulk, and reaches that material's bulk $B_{sh}$ while $B_i$ is still relatively small. However, there is a small range of film thicknesses where the film and the bulk can simultaneously be close to their respective $B_{sh}$. In this case, the $B_{sh}$ of the SIS structures is somewhat higher than the bulk value of the film material, but the potential gain in $B_{sh}$ is relatively small, and the gain decreases as the thickness of the insulating layer increases. The $B_{sh}$ gain can be explained with a rough argument invoking the superfluid velocity $v_s$:
\begin{equation}
\bv_s  = \frac{\hbar}{m^*}\left(\nabla \phi - \frac{e^*}{\hbar}\bA\right)
\label{eq:vs}
\end{equation}
where $m^*$ and $e^*$ are the effective cooper pair mass and charge, $\bA$ is the magnetic vector potential, and $\nabla \phi$ is the gradient in the phase of the order parameter. If the superconductor is in the Meissner state, $\nabla\phi=0$ and the magnitude of $\bA$ can be approximated by $A=\int{B\mathrm{d}x}$, where the integration starts deep in the bulk superconductor where $A=0$ and proceeds outwards. If the bulk superconductor is a material with a relatively small $\lambda$, such as niobium, then at the film $A$ and therefore $v_s$ will be smaller than if the bulk material were a large-$\lambda$ material, such as the alternative materials under consideration. As a result, the superconductors in the SIS structure are relatively ``unstressed,'' and can screen a larger external field. This same argument explains why it is optimal for $\delta$ to be as small as possible---this keeps $A$ as small as possible. However, the gain in $B_{sh}$ cannot be multiplied by adding more films of the same material. As with the homolaminate, $B_M$ (and the vector potential) will be unchanged by splitting up the superconducting film into separate layers with thin insulators between them, but $B_V$ will have a stronger influence on the vortex, pulling it into the film. This detrimental effect is visible in the last line of Figure \ref{fig:BshvsdMulti}, showing the calculation for a series of 5 Nb$_3$Sn films, each with thickness $d/5$, on a Nb bulk, with thin insulating films between the superconductors.

Note that for the material parameters used, which came from experimental references, the theoretical superheating fields of NbN and MgB$_2$ are smaller than that of Nb. In this study, calculations were not performed for the SS' structure, a superconducting film deposited directly onto a different superconductor, without an insulating layer. The calculation would be significantly more difficult, but such a structure may offer a stronger performance improvement for appropriate choices of the thickness of the film layer and the superconducting materials.

\section{Dissipation Above Bsh}
\label{sec:dissipation}

Gurevich states \cite{Gurevich} that ``thin film coating significantly decreases vortex dissipation at $B_0>B_v$,'' where $B_v$ is the vortex penetration field. This raises the question of whether it might be possible to use a SIS structure above $B_v$, in a regime in which vortices pass through the film each half cycle, bringing trapped flux into and out of the insulating region. Gurevich gives an expression for the power dissipated per area:
\begin{equation}
P/A = \frac{2 \omega d}{\pi \mu_0 \lambda_f}\left(\lambda_b + \delta + d/2\right)B_v\left(B_0-B_v\right)
\label{eq:Pdiss}
\end{equation}
where $B_v\approx B_{sh}$. Consider a $d=50$ nm Nb$_3$Sn film in a SIS structure with a Nb$_3$Sn bulk and a very thin insulator, exposed to a field $B_0$ that is 1 mT higher than the film's $B_{sh}$. In this case, $P/A \approx 9$ W/cm$^2$. For a single cell TESLA cavity, even this very small excess over $B_{sh}$ generates approximately 4 kW of heat, far too much to be feasible for SRF applications.

\section{Shielding in DC}
\label{sec:DC}

As Eqn.~\ref{eq:Pdiss} shows, the power dissipated by vortex penetration is proportional to frequency. For low frequency AC and DC applications, vortex penetration could be tolerated. In this case, SIS structures may offer a distinct advantage. As Eqn.~\ref{eq:vs} shows, the superfluid velocity, a measure of the ``stress'' in a superconductor, is reduced in a large $A$ field by a gradient in the phase of the order parameter $\nabla\phi$. Even though the $A$ field at the film is tied to that of the bulk (since the $A$ field is continuous across the insulating gap), the insulating gap does allow the opportunity to decouple $\nabla\phi$ in the film from that in the bulk. This is accomplished by passing vortex lines through the superconducting film, which would incur large dissipation at high frequencies, but is permissable at low frequencies. With several layers, and an appropriate $\nabla\phi$ maintained in each of them to compensate for the $A$ field, it should be possible to screen even very large fields while maintaining a relatively small $v_s$ in each of the layers, so that they remain in the Meissner state.

\section{Performance Above Bc1 in Bulk Films}
\label{sec:bulk}

The primary motivation for trying to fabricate SIS structures was the promise of enhancing the $B_{c1}$ of low-$\xi$ alternative materials. It was suggested that above $B_{c1}$, dissipation would occur due to vortex dissipation \cite{Gurevich}. There was some experimental data supporting this idea, as Nb$_3$Sn cavities coated by Wuppertal researchers showed increasing surface resistance with field, an effect that onset near the $B_{c1}$ of Nb$_3$Sn \cite{wup}\cite{Hasan}. However, it was not clear if this performance degradation was fundamental and potentially related to vortex dissipation or if it was related to some other loss mechanism that could be ameliorated.

This year, a Nb$_3$Sn cavity was coated and tested at Cornell. Vertical test data and fits to material parameters show that the cavity clearly exceeds $B_{c1}$ without a signficant increase in surface resistance \cite{Sam}\cite{Matthias}. The $B_{c1}$ value calculated from the material parameters was confirmed by $\mu$-SR measurements on a witness sample \cite{Anna}. Furthermore, a niobium cavity was prepared and tested at Cornell after receiving a furnace treatment that gave it a very small mean free path. As a result, the niobium had a very small $\xi$, similar to that of the alternative materials under consideration. It too had a performance with minimal increase in surface resistance, reaching fields significantly higher than the $B_{c1}$ value determined from fits to material parameters \cite{Dan}\cite{footbc1}.

\section{Discussion and Conclusions}
\label{sec:Conclusions}

In this paper, we have shown that contrary to suggestions that SIS structures enhance $B_{c1}$, in fact they reduce it to zero. In addition, it was shown that the $B_{sh}$ of an SIS structure is only marginally larger than the bulk value and only for a small parameter space, and that using a multilayer only decreases $B_{sh}$ of the film. It was also shown that SIS structures exhibit unmanageable levels of heating above $B_{sh}$ at high frequencies. Therefore, it seems that SIS structures are not beneficial for SRF applications. However, they may be useful in DC and low frequency applications, where it should be possible to set up a gradient in the phase of the order parameter in the thin films, allowing them to screen very large fields. 

Based on the results of this study, the authors of this paper recommend that SRF researchers developing alternative materials concentrate their efforts on bulk films. Bulk films are quite simple to fabricate compared to SIS films, but they offer a similar ideal SRF performance. And although we have shown that it is not possible to augment $B_{c1}$ with SIS structures, there is still great promise for alternative materials. Because of the recent experiments showing that low-surface-resistance operation above $B_{c1}$ is possible with cavities made from short coherence lengths superconductors, we now know that the potential of bulk films has not yet been realized.


\begin{thebibliography}{99}
\bibitem{footgibbs}
One definition of $B_{c1}$ might be the field at which there is a stable position for a vortex inside a superconductor. However, in our definition, the stable position for the flux might be inside the insulating region, in which case there will be no vortex, as a vortex has a normal conducting core in a superconductor. It will be shown in this paper that our definition is the relevant one, because if vortices move from the external region across the superconductor to become trapped flux in the insulating region, a very large amount of heating will occur. In other words, above $B_{c1}$ as we have defined it, the film is in a metastable state, and only an energy barrier prevents strongly dissipative flux penetration.
\bibitem{foot2}
These calculations use the London model, valid only for strongly type II materials. It neglects non-linearity, so at high fields, it is only valid qualitatively, and likely overestimates $B_{sh}$. In fact, $B_{sh}$ calculated in this paper is the field at which the vortex penetration barrier disappears, an upper bound to the actual $B_{sh}$. The latter is calculated by a linear stability analysis, similar to that done in \cite{Transtrum} for bulk superconductors, that will be presented elsewhere. However the $B_{sh}$ found in this paper does seem to be a good approximation: the $B_{sh}$ for bulk Nb$_3$Sn in Fig. \ref{fig:Gplots} is very close to the calculated value from Tab. \ref{tab:mats}.
\bibitem{foot1}
A more basic version of these $B_{sh}$ calculations of the SIS structure have been performed previously by Kubo, Iwashita, and Saeki \cite{Kubo}.
\bibitem{footbc1}
For calculations of $B_{c1}$ from material parameters, Ginzburg Landau theory is applied outside its validity, so corrections on the order of a few percent will apply. For example, using GL theory at low temperatures is known to underestimate the bulk superheating ﬁeld (see Catelani and Sethna, PRB 78.224509).
\end{thebibliography}

\end{document}